\documentclass[aps,prl,twocolumn,groupedaddress,floatfix,showpacs]{revtex4}
\usepackage{graphicx}
\usepackage{amsmath}
\def\sbf#1{\mbox{\boldmath $#1$}}
\begin{document}

\title{Andreev interference in adiabatic pumping}

\author{Fabio Taddei, Michele Governale, and Rosario Fazio}

\affiliation{
NEST-INFM \& Scuola Normale Superiore, I-56126 Pisa, Italy}

\date{\today}
\begin{abstract}
Within the scattering approach, we develop a model for adiabatic
quantum pumping in hybrid normal/superconductor systems where several
superconducting leads are present.
This is exploited to study Andreev-interference effects on
adiabatically pumped charge in a 3-arm beam splitter attached to one normal
and two superconducting leads with different phases of the order parameters.
We derive expressions for the pumped charge through the normal lead
for different parameters for the scattering region, and elucidate the effects due to
Andreev interference.
In contrast to what happens for voltage-driven transport,
Andreev interference does not yield in general a
pumped current which is a symmetric function of 
the superconducting-phase difference.
\end{abstract}
\pacs{ 73.23.-b, 74.45.+c}
\maketitle

\textit{Introduction}. Pumping consists in the transport of particles obtained, 
in absence of a transport voltage, 
by varying in time some properties of a mesoscopic conductor. 
If the time scale for the variation of the scattering 
matrix describing the conductor is larger than the transport time, 
then the pumping is \emph{
adiabatic} and the number of particles 
transferred per period does not depend on the detailed 
time evolution of the scattering matrix  but only on geometrical 
properties of the pumping cycle~\cite{brouwer}. 

Adiabatic pumping has attracted a vast interest, and 
different aspects of this phenomenon  have been addressed
\cite{makhlin,levitov,levinson,aleiner,moskalets,butti} as, for example, 
the counting statistics of the 
pumped current, the generalization to multi-terminal geometries and 
the question of phase coherence. 
The idea of adiabatic pumping has been combined with other phenomena 
typical of  mesoscopic physics, like spin-dependent 
transport\cite{mucciolo,spinpump,watson}, 
Kondo physics\cite{wangpumping,aono}, Luttinger-liquid physics\cite{citro}, 
and  Quantum Hall effect\cite{quantumhall}. 
So far, there have been only few investigations of adiabatic pumping
in normal/superconductor hybrid structures. 
Zhou\cite{zhoucondmat} has considered the pumped current due to the 
time-modulation 
of the superconducting correlations induced in the normal region. 
Wang \textit{et al.}\cite{wangapl} 
have studied the combined effect of pumping and Andreev 
reflection in a system with only one single-mode superconducting lead,  
finding up to a fourfold enhancement 
of the pumped current due to the interplay of Andreev and normal reflection. 
The generalization to a multi-mode superconducting lead was done 
by Blaauboer\cite{blaauboer}. 

In this paper we explore the physics of adiabatic pumping in 
the presence of several superconducting leads. 
In particular, we want to study Andreev interference in 
adiabatic pumping. 
Andreev interferometers have been intensively investigated in the past both
in the diffusive limit
\cite{hekking_93,hui_93,nazarov_94,zaitsev_94,nazarov_96,volkov_96}, 
and in  the ballistic one\cite{nakano,takagi} (for an extended list of
references see, for example, Ref. \cite{lambert_98}).
In a standard Andreev 
interferometer (as those considered in Refs. 
\cite{hekking_93,nazarov_94,nazarov_96,volkov_96,nakano,takagi}) 
transport is driven by an applied voltage. In the present work, 
we study the problem of an Andreev interferometer when transport is  
induced  by adiabatic pumping. 

The paper is organized as follows: we start by deriving a 
formula for the current pumped through a normal lead in the presence 
of several other normal and superconducting leads; then we apply the 
formalism to a fork-shaped structure 
which exhibits Andreev interference. 

\textit{Formalism.} 
We consider a system consisting of $P$ normal and $Q$ superconducting 
leads connected to a generic scattering region
characterized by its scattering matrix $\sbf{\mathcal{S}}$ (matrices are in boldface).
The different 
superconductors are described by a constant  pair potential 
$\Delta_m= |\Delta_m| \exp {(i \Phi_m)}$, where $m=1,...,Q$ labels the 
superconducting leads.
We note that 
when all leads are in the normal state 
we can write 
the scattering matrix as
\begin{equation} 
\label{se}
\sbf{S}_{\text{N}}=\left(\begin{array}{cc}
\sbf{R}(\epsilon) & \sbf{T}^{\prime}(\epsilon)\\
\sbf{T}(\epsilon) &  \sbf{R}^{\prime}(\epsilon)
\end{array}
\right),
\end{equation}
where $\sbf{R}$ is a $P \times P$ matrix containing 
the scattering amplitudes between the normal terminals; 
$\sbf{R}^{\prime}$ is a $Q \times Q$ matrix containing 
the scattering amplitudes between the superconducting terminals;
$\sbf{T}$ is a $Q \times P$ matrix describing the scattering 
between the normal leads and the superconducting ones; and 
$\sbf{T}^{\prime}$ 
is $P \times Q$ matrix describing the scattering between the 
superconducting leads and the normal ones. 
The energy $\epsilon$ is measured with respect to the Fermi energy. 
Writing the scattering matrix as in Eq. (\ref{se}) makes evident 
that the system is equivalent to one consisting only of a 
normal lead with $P$ modes and a superconducting lead with $Q$ modes, 
each mode in the superconducting lead having its own pair potential  
$\Delta_{m}$. 
We can now write the scattering matrix for the hybrid normal--superconductor 
system in Nambu space:
\begin{equation}
\label{matrixr}
\sbf{\mathcal{S}}=\left(\begin{array}{cc}
\sbf{\mathcal{R}}_{\text{ee}}(\epsilon) & -{\sbf{\mathcal{R}}_{\text{he}}}^{*}(-\epsilon)\\
\sbf{\mathcal{R}}_{\text{he}}(\epsilon) & {\sbf{\mathcal{R}}_{\text{ee}}}^{*}(-\epsilon)
\end{array}
\right),
\end{equation}
where the submatrices are obtained composing the matrix
$\sbf{S}_{\text{N}}$ with the scattering matrix $\sbf{S}_{\text{NS}}$ of a perfect NS 
multichannel interface~\cite{beenakker}.
The latter is a diagonal matrix of Andreev reflection amplitudes which 
can be written,
under the Andreev approximation, as
$\sbf{S}_{\text{NS}}=\sbf{\alpha} ~\sbf{e^{-i \Phi}}$, where
$\sbf{\alpha}$ is a diagonal matrix whose 
elements are $\exp[-i \text{Arcos}(\epsilon/|\Delta_{m}|)]$ and
$\sbf{e^{-i \Phi}}$ is a diagonal matrix of elements $\exp[-i \Phi_m]$.
As a result:
\begin{subequations}
\label{relem}
\begin{eqnarray}
\mathbf{\sbf{\mathcal{R}}_{\text{ee}}}(\epsilon)&=& \sbf{R}(\epsilon)+\nonumber\\
& &
\sbf{T}^{\prime}(\epsilon) \sbf{\alpha e^{i \Phi}} {\sbf{R}^{\prime}}^{*}(-\epsilon) 
\sbf{\alpha} \sbf{e^{-i \Phi}}
\sbf{M}(\epsilon)\sbf{T}(\epsilon)\\
\sbf{\mathcal{R}}_{\text{he}}(\epsilon)&=&
{\sbf{T}^{\prime}}^{*}(-\epsilon) \sbf{\alpha} \sbf{e^{-i \Phi}}
\sbf{M}(\epsilon)\sbf{T}(\epsilon)
\end{eqnarray}
\end{subequations}
with
\begin{equation}
\label{me}
\sbf{M}(\epsilon)=\left[\mathbf{1}-\sbf{R}^{\prime}(\epsilon)\sbf{\alpha e^{i \Phi}} 
{\sbf{R}^{\prime}}^{*}(-\epsilon) 
\sbf{\alpha e^{-i \Phi}}
\right]^{-1}.
\end{equation} 
In Eq. (\ref{matrixr}), $\sbf{\mathcal{R}}_{\text{ee}}$ ($\sbf{\mathcal{R}}_{\text{he}}$) is a
$P\times P$ matrix of normal (Andreev) scattering amplitudes between the
normal leads.
Note that we have used the particle-hole symmetry, which yields
$\sbf{\mathcal{R}}_{\text{hh}}(\epsilon)=\sbf{\mathcal{R}}^*_{\text{ee}}(-\epsilon)$ and
$\sbf{\mathcal{R}}_{\text{eh}}(\epsilon)=-\sbf{\mathcal{R}}^*_{\text{he}}(-\epsilon)$.

By means of the scattering matrix Eq.~(\ref{matrixr}), 
operating along the same lines of Refs.~\cite{blaauboer,buettiker_02}, 
we can write the charge pumped through any of the normal leads: 
\begin{equation}
\label{pumpedq}
Q_n=\frac{e}{\pi}\int dX_1 dX_2 
\sum_{l\in \text{Normal leads}}  \Pi_{n,l} (X_1,X_2),
\end{equation}
where $X_1$, and $X_2$ are the two pumping fields,  
and 
\begin{widetext}
\begin{equation}
\label{pinl}
\Pi_{n,l} (X_1,X_2)=\int_{-\infty}^\infty d\epsilon \left(-
\frac{\partial f }{\partial \epsilon}\right)\mbox{Im} \left\{
\frac{\partial\left({\sbf{\mathcal{R}}_{\text{ee}}}^{*}(\epsilon)\right)_{n,l} }
{\partial X_1}\frac{\partial\left({\sbf{\mathcal{R}}_{\text{ee}}(\epsilon)}\right)_{n,l} }
{\partial X_2}  
-\frac{\partial\left( {\sbf{\mathcal{R}}_{\text{he}}}^{*}(\epsilon)\right)_{n,l} }
{\partial X_1}\frac{\partial\left(\sbf{\mathcal{R}}_{\text{he}}(\epsilon)\right)_{n,l}}
{\partial X_2}
\right\}, 
\end{equation}
\end{widetext}
with $f(\epsilon)$ being the Fermi Dirac distribution.  
Although Eq.~(\ref{pinl}) is valid at finite temperature, in the 
rest of the paper we will restrict ourselves to zero temperature.  

\begin{figure}[b]
\includegraphics[width=3.in]{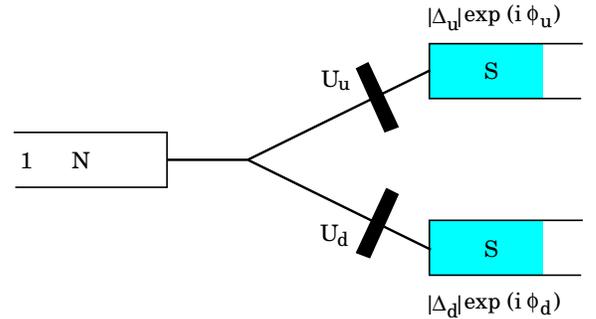}
\caption{ Schematic picture of the Andreev interferometer, consisting of 
a symmetric beam splitter with $\delta$-barriers added on the two arms 
where the superconducting leads are connected. The strength of the 
$\delta$-barriers can be varied in time, and are used as pumping fields.  
\label{interfer}}
\vspace{1cm}
\end{figure}

\textit{Andreev interferometers.}
We, now, apply the formalism developed above 
to an example of an Andreev interferometer.
The most simple system which allows us to investigate pumping with different
superconductors consists of a 3-arm beam splitter schematically 
shown in Fig. \ref{interfer}. For the sake of simplicity we consider 
a symmetric beam splitter, 
whose scattering matrix (in the normal state) is  
\begin{equation}
\label{fork}
\sbf{S}_{\text{B}}=\left(
\begin{array}{ccc}
  -s_1\sqrt{1-2\gamma} & \sqrt{\gamma} & \sqrt{\gamma} \\
\sqrt{\gamma}& a & b \\
\sqrt{\gamma}& b & a\\
\end{array}
\right)
\end{equation}
where $a=1/2(s_2+s_1\sqrt{1-2\gamma})$, $b=1/2(-s_2+s_1\sqrt{1-2\gamma})$,
with $s_i=\pm 1$ and $0\le\gamma\le 1/2$.
Lead 1, 
on the left-hand-side of Fig.~\ref{interfer}, is normal metallic,
while the other two leads, 
denoted by u and d, are superconducting with order parameters, respectively,
equal to $|\Delta_{\text{u}}|\exp{(i\Phi_{\text{u}})}$ and $|\Delta_{\text{d}}|
\exp{(i\Phi_{\text{d}})}$.

The parameters to be varied in time are the  strengths,
$U_{\text{u}}$ and $U_{\text{d}}$, of two additional 
$\delta$-barriers placed 
in the two arms on the right side of Fig.~\ref{interfer}.
When all leads are in the normal state, the total scattering matrix 
$\sbf{S}_{\text{N}}$ is obtained
by combining the scattering matrix $\sbf{S}_{\text{B}}$ given in 
Eq.~(\ref{fork}) 
with the transmission and reflection 
amplitudes of the two $\delta$-barriers, namely
\begin{equation}
t =\frac{1}{1+i \beta_{\text{u/d}}}
\end{equation}
and
\begin{equation}
r=-i \frac{\beta_{\text{u/d}}}{1+i \beta_{\text{u/d}}}
\end{equation}
where 
$\beta_{\text{u/d}}=U_{\text{u/d}}/(\hbar v_{\text{F}})$, $v_{\text{F}}$ 
being the Fermi velocity.
We choose as pumping fields $\beta_{\text{u/d}}$, i.e. the strength of 
the $\delta$-barriers. We consider 
the sinusoidal week pumping 
limit: $\beta_{\text{u}}=\bar{\beta}_{\text{u}}+\delta\beta_{\text{u}} 
\sin(\omega t)$ and $\beta_{\text{d}}=\bar{\beta}_{\text{d}}+\delta
\beta_{\text{d}} \sin(\omega t -\delta\phi)$.

First we start by studying the case when the normal lead is tunnel-coupled to 
the rest of the structure, obtained by setting $\gamma\rightarrow 0$, $s_1=-1$ and $s_2=1$ in Eq.~(\ref{fork}). 
The analytical expression for the charge pumped through lead 1, in 
leading order in $\gamma$, reads 
\begin{equation}
\label{qstunn}
Q_{1}^{\text{S}}=- A \frac{e}{\pi} \gamma^2 4 \sin(\Delta \Phi)\frac{ \bar{\beta}_{\text{u}}^2+ 
 \bar{\beta}_{\text{d}}^2 -2 \bar{\beta}_{\text{u}}\bar{\beta}_{\text{d}} 
\cos(\Delta\Phi)}{\left[\left(\bar{\beta}_{\text{u}}+ 
 \bar{\beta}_{\text{d}}\right)^2 +
\cos^2\left(\frac{\Delta\Phi}{2}\right)\right]^3},
\end{equation} 
being $\Delta \Phi=\Phi_{\text{u}}- \Phi_{\text{d}}$ the phase 
difference, and $A=\pi \delta
\beta_{\text{u}}\delta
\beta_{\text{d}}\sin\delta\phi$ the area of the pumping cycle in parameter 
space.
The $\gamma^2$ dependence of the charge in Eq.~(\ref{qstunn}) is the 
expected one for transport mediated by  Andreev reflection. 
It is interesting to note that  $Q_1^\text{S}$ is an odd function of the 
phase difference, and that no pumping occurs at zero phase ($\Delta\Phi=0$). 
The pumped charge $Q_1^\text{S}$ can be contrasted with the linear DC conductance 
$G_{\text{DC}}=\frac{e^2}{h} 2 |\left(\sbf{\mathcal{R}}_{\text{he}}\right)_{1,1}|^2$
of the system \cite{lambert_98}, with the barriers frozen at their average values 
$\bar{\beta}_{\text{u,d}}$, when a transport voltage is applied between
the normal and the superconducting terminals (superconductors being at the same
potential).  
For this particular case, in leading order in $\gamma$,  $G_{\text{DC}}$ 
reads
\begin{widetext}
\begin{equation}
\label{dctunn}
G_{\text{DC}}=\frac{e^2}{h} \gamma^2 \frac{1+4\left(
\bar{\beta}_{\text{u}}^2+\bar{\beta}_{\text{d}}^2\right)+8\left(
\bar{\beta}_{\text{u}}^4+\bar{\beta}_{\text{d}}^4\right)+(1+4\bar{\beta}_{\text{u}}^2)(1+4\bar{\beta}_{\text{d}}^2)\cos(\Delta\Phi)}
{\left[\left(\bar{\beta}_{\text{u}}+ 
 \bar{\beta}_{\text{d}}\right)^2 +
\cos^2\left(\frac{\Delta\Phi}{2}\right)\right]^2}. 
\end{equation}
\end{widetext}
The DC conductance, Eq.~(\ref{dctunn}), is an even function of the phase 
difference. It has a zero-phase extremum, which can be either a maximum or 
minimum depending on the strength of the barriers. 
To complete the analysis, we mention that the system 
acts as a pump also when all leads are in the normal state. To
leading order the pumped charge is linear in $\gamma $ 
(as expected for vanishing superconducting gap), 
and it reads
\begin{equation}
Q_{1}^{\text{N}}=
- A \frac{e}{\pi} 4 \gamma\frac{\bar{\beta}_{\text{u}}-\bar{\beta}_{\text{d}}}
{\left[\left(\bar{\beta}_{\text{u}}+\bar{\beta}_{\text{d}}\right)^2+1\right]^2}
\quad.
\end{equation}
In contrast to the superconducting case, the leading order of the pumped charge vanishes
when $\bar{\beta}_{\text{u}}=\bar{\beta}_{\text{d}}$.

Now, let us turn to the case of a  maximally-transmitting beam splitter, which is
obtained from Eq.~(\ref{fork}) setting $\gamma=1/2$ and $s_{1,2}=1$.
The analytical form for the charge pumped through lead 1 is rather involved
and we report only the limits $\bar{\beta}_{\text{u/d}} \ll1$ (large barrier transmission)
and  $\bar{\beta}_{\text{u/d}} \gg 1$ (small barrier transmission):
\begin{equation}
\label{q1smax}
Q_{1}^{\text{S}}=\left\{
\begin{array}{ll}\displaystyle
256 A \frac{e}{\pi}\frac{( \bar{\beta}_{\text{u}}- \bar{\beta}_{\text{d}}) 
\cos^2\left
(\frac{\Delta\Phi}{2}\right)}{\left[3+\cos(\Delta\Phi)\right]^3} & \mbox{if}\quad 
\bar{\beta}_{\text{u/d}} \ll1\\ \\ \displaystyle
-A \frac{e}{\pi}\frac{\sin(\Delta\Phi)}
{4 \bar{\beta}_{\text{u}}^3 \bar{\beta}_{\text{d}}^3} &  \mbox{if}\quad 
\bar{\beta}_{\text{u/d}} \gg 1
\end{array}\right. .
\end{equation}
It is interesting to note that while in the case of large barrier transmission $Q_1^\text{S}$ is an even function of the phase difference, for small transmission $Q_1^\text{S}$
is odd. 
However no definite parity is present for arbitrary transmissions $\bar{\beta}_{\text{u/d}}$ and, in
particular, when $\bar{\beta}_{\text{d}}=0$ we obtain:
\begin{equation}
Q_{1}^{\text{S}}=\frac{128\bar{\beta}_{\text{u}}[1+(1+
4{\bar{\beta}_{\text{u}}}^2)
\cos \Delta\phi-{\bar{\beta}_{\text{u}}}\sin \Delta\phi ]}
{[3+10{\bar{\beta}_{\text{u}}}^2+\cos \Delta\phi]^3} ~.
\end{equation}
Again, it is instructive to see what happens for the DC conductance, when the 
barriers are frozen to their average value. Also for this case, we report 
the same two limiting cases shown for $Q_{1}^{\text{S}}$
 \begin{equation}
G_{\text{DC}}=\left\{
\begin{array}{ll}\displaystyle
\frac{e^2}{h} 32 \frac{\cos^2\left(\frac{\Delta \Phi}{2}\right)}
{\left[3+\cos(\Delta \Phi)\right]^2}
 & \mbox{if}\quad 
\bar{\beta}_{\text{u/d}} \ll1\\ \\ \displaystyle
\frac{e^2}{h} \frac{ \bar{\beta}_{\text{u}}^4+\bar{\beta}_{\text{d}}^4
+2 \bar{\beta}_{\text{u}} \bar{\beta}_{\text{d}} \cos(\Delta \Phi)}
{8\bar{\beta}_{\text{u}}^4\bar{\beta}_{\text{d}}^4} &  \mbox{if}\quad 
\bar{\beta}_{\text{u/d}} \gg 1
\end{array}\right. .
\end{equation} 
The DC conductance is an even function of the phase difference  both
large and small barrier transmission.
It, actually, remains even also 
for arbitrary values of $\bar{\beta}_{\text{u/d}}$, while 
its zero-phase extremum can be either 
a maximum or a minimum depending on the values of $\bar{\beta}_{\text{u/d}}$. 
Finally, we report the pumped charge when the system is in the normal 
state
\begin{equation}
\label{q1nmax}
Q_{1}^{\text{N}}=\left\{
\begin{array}{ll}\displaystyle
- A \frac{e}{\pi}(\bar{\beta}_{\text{u}}-\bar{\beta}_{\text{d}}) & \mbox{if}\quad 
\bar{\beta}_{\text{u/d}} \ll1\\ \\ \displaystyle
-A \frac{e}{\pi}\frac{1}{8 \bar{\beta}_{\text{u}}^2 \bar{\beta}_{\text{d}}^2 }
\left(\frac{1}{ \bar{\beta}_{\text{u}}}-\frac{1}{ \bar{\beta}_{\text{d}}}\right) &  \mbox{if}\quad 
\bar{\beta}_{\text{u/d}} \gg 1
\end{array}\right. .
\end{equation}
Contrasting Eq.~(\ref{q1smax}) with Eq. (\ref{q1nmax}), we notice 
that for the case $\bar{\beta}_{\text{u/d}} \ll1$ ,  
to leading order in the pumping parameters, superconductivity produces
an enhancement of the charge pumped, reaching a maximum of a factor 4.

Finally, we wish to point out that the most distinctive signature of 
Andreev interference in the adiabatic pumping regime is the lack of a 
definite symmetry of the pumped current 
under inversion of the superconducting-phase difference. On the contrary,  
the DC current produced by an applied transport voltage, either DC or 
time-dependent,
is always an even function of $\Delta \Phi$. 
The case of a DC voltage has been discussed above.
It can be easily seen that
also the current produced 
by rectification of an oscillating voltage $V_{\text{osc}}$ 
(for example induced by the pumping voltages on stray capacitances
\cite{rectification}) is 
an even function of $\Delta \Phi$.
In fact, the current produced by rectification reads 
\cite{rectification,moskalets_01}  
$I_{\text{rect}}=
\omega/2\pi\int_0^{2\pi/\omega}G(\tau) V_{\text{osc}}(\tau) d\tau$, 
where $G(\tau)$ is the instantaneous conductance 
which is an even function of $\Delta\Phi$, and all other quantities do not depend 
on the superconducting-phase difference.
This lack of symmetry with respect to superconducting-phase difference
can be exploited to distinguish between pumping and rectification.
This is analogous to the normal case where the symmetry used for this purpose is 
the one related to magnetic field inversion \cite{rectification}. 

\textit{Conclusions.}
In this paper we have derived a scattering formula for adiabatically pumped
charge in hybrid NS multi-terminal systems.
This has been used to study Andreev interference in a 3-arm beam splitter
attached to one normal and two superconducting leads with different phases of the
order parameters.
Within the weak pumping limit we found that Andreev interference very
much affects the charge pumped through the normal lead, though differently
with respect to the case of DC-voltage-driven transport.
In general, the pumped charge has no definite symmetry under inversion of the
superconducting-phase difference and no zero-phase extremum is found.

\end{document}